\documentclass[prb,twocolumn,a4paper]{revtex4}
\usepackage{graphicx}
\usepackage{wrapfig}
\usepackage{epstopdf}
\usepackage{float}
\usepackage{bm}
\usepackage[latin1]{inputenc}
\usepackage{placeins}

\begin{document}

\title{GaAs Quantum Dot Thermometry Using Direct Transport and Charge Sensing}

\author{D.~Maradan}
\thanks{These authors contributed equally to this work.}
\email{d.maradan@unibas.ch}
\author{L.~Casparis} 
\thanks{These authors contributed equally to this work.} 
\email{lucas.casparis@unibas.ch}
\author{T.-M.~Liu} 
\author{D.~E.~F.~Biesinger} 
\author{C.~P.~Scheller} 
\author{D.~M.~Zumb\"uhl} 
\email{dominik.zumbuhl@unibas.ch} 
\affiliation{Department of Physics, University of Basel, Klingelbergstrasse 82, 4056 Basel, Switzerland}
\author{J.~Zimmerman}
\author{A.~C.~Gossard}
\affiliation{Materials Department, University of California, Santa Barbara, California, USA}

\begin{abstract}
We present measurements of the electron temperature using gate defined quantum dots formed in a GaAs 2D electron gas in both direct transport and charge sensing mode. Decent agreement with the refrigerator temperature was observed over a broad range of temperatures down to $10\,$mK. Upon cooling nuclear demagnetization stages integrated into the sample wires below $1\,$mK, the device electron temperature saturates, remaining close to $10\,$mK. The extreme sensitivity of the thermometer to its environment as well as electronic noise complicates temperature measurements but could potentially provide further insight into the device characteristics. We discuss thermal coupling mechanisms, address possible reasons for the temperature saturation and delineate the prospects of further reducing the device electron temperature.
\end{abstract}
\maketitle

\section{Introduction}
\label{introduction}

Two-dimensional electron gases (2DEGs) are a versatile, widely-used experimental platform in low temperature solid state physics because of their nearly ideal two-dimensional nature and their possibility to confine electrons to almost arbitrary shapes using gate voltages. Groundbreaking experiments have been realized in these systems, including artificial atoms~\cite{vanhouten_1992,Kastner_1992,kouwenhoven_1996}, the integer and fractional quantum hall effect~\cite{klitzing_1980,tsui_1982} and spin qubits~\cite{loss_1998,hanson_2007}. In many experiments, the temperature of the 2DEG is much higher than the temperature $T_{\rm MC}$ of the dilution refrigerator mixing chamber due to various reasons, including poor thermal coupling and insufficient filtering. However, a wide range of phenomena contain small energy scales and are only accessible at very low temperatures. These include nuclear spin physics~\cite{simon_2007,simon_2008}, novel quantum phases~\cite{scheller_2013} and multiple impurity~\cite{jayaprakash_1981} or multiple channel~\cite{nozieres_1980,zawadowski_1980} Kondo physics. Further, studies of fragile fractional quantum Hall states, including candidates for non-Abelian physics such as the $\nu = 5/2$ state~\cite{willett_1987}, would benefit from low temperatures, possibly opening the doors for topological quantum computation~\cite{nayak_2008}.

To our knowledge, the lowest reliable temperature reported in a 2DEG is $4\,$mK~\cite{pan_1999,Samk_2011} in a fractional quantum Hall experiment, with sintered silver heat exchangers attached to the sample wires in a \textsuperscript{3}He cell. In Ref.\,\cite{pan_1999}, a PrNi$_5$ demagnetization stage at $0.5\,$mK was used to cool the liquid $^3$He, well below the $4\,$mK of the 2DEG sample. For quantum Hall samples loaded into a chip holder in vacuum, slightly higher temperatures $9\ldots 13\,$mK were reported \cite{Chung_2003,Radu_2008}. Interestingly, in Ref.\,\cite{Radu_2008} (supplementary), the refrigerator base temperature was below $6\,$mK and the temperature measured with a Coulomb blockaded quantum dot was $16\pm3\,$mK. The lowest GaAs quantum dot temperature measurement reported is 12~mK~\cite{potok_2007,Karakurt_2001}, as far as we know.

These examples indicate that cooling of a 2DEG embedded in a semiconductor such as e.g. GaAs is a difficult task. The main reason is the weakening of the electron-phonon interaction ($\propto T^5$) at low temperatures. Therefore, at very low temperature, the system benefits from cooling through the conduction electrons (Wiedemann-Franz mechanism, $\propto T^2$), where heat transfer is mediated through the electrical contact to the sample. For typical semiconductor devices with large contact resistances, this comparably weak coupling makes the sample vulnerable to heat leaks, e.g. high frequency radiation or dissipative heating. Additionally, the weakening of the electron-phonon interaction significantly complicates the thermal coupling of the insulated sample wires to the coldest part of the refrigerator.

Recently, we have proposed a way to overcome these limitations by integrating a copper nuclear refrigerator into each of the electrical sample wires connected to an electronic transport sample, providing efficient thermal contact to a bath at low mK or microkelvin temperature \cite{clark_2010}. For efficient precooling of the nuclear refrigerators as well as for regular dilution refrigerator operation, every sample wire is connected to a sintered silver heat exchanger located in the plastic mixing chamber of the dilution refrigerator with a base temperature of $9$~mK. Further, to minimize the effect of high-frequency radiation, all electrical lines are filtered extensively using thermocoax cables, cryogenic Ag-epoxy microwave filters~\cite{scheller} and double-stage RC filters of bandwidth $30\,$kHz. The measurement setup is described in detail in reference~\cite{casparis_2012}. In semiconductor samples such as GaAs 2DEGs, the ohmic contacts will probably present the largest electrical and thermal impedance in this cooling scheme.

\section{Quantum Dot Thermometry}
\label{qdelectronthermometer}

Gate defined GaAs quantum dots in deep Coulomb blockade are used as a thermometer directly probing the electron temperature $T$ in the surrounding 2DEG by measuring the thermal smearing of the Fermi edge~\cite{vanhouten_1992}. The quantum dot is coupled to two electron reservoirs via left and right tunnel barriers with tunnel rates $\Gamma_L$ and $\Gamma_R$. In the symmetric case $\Gamma_L = \Gamma_R = \Gamma$, the direct current through the quantum dot is approximated by $I_{\rm DC} = e\Gamma/2$ assuming sequential tunneling, with $e$ the electron charge. In the temperature broadened Coulomb blockade regime ($h \Gamma \ll k_B T $, with Boltzmann constant~$k_B$ and Planck constant~$h$), the narrow dot level with broadening $\sim \Gamma$ acts as a variable energy spectrometer which can resolve and directly map the Fermi-Dirac (FD) distribution in the current through the dot. The energy of the spectrometer can be tuned by capacitively shifting the dot energy level with a gate, e.g. the plunger gate at voltage $V_P$. With a sufficiently large DC source-drain bias $V_{\rm SD}\gg k_B T/e$, the chemical potential of source and drain reservoirs can be individually resolved, separately giving the distribution functions of each reservoir when sweeping the plunger gate voltage $V_P$ through both source and drain chemical potentials.

To stay in the single level transport regime, the bias $V_{\rm SD}$ has to be small compared to the excited state energy $\Delta$. To obtain the temperature from each distribution function, the gate lever arm $\alpha$ is required for the conversion from gate voltage to energy. The separation $\Delta V_P$ in gate voltage between the inflection points of the two FD distributions can be taken from the plunger gate sweep $I_{\rm DC}(V_P)$ at a fixed, known bias $V_{\rm SD}$. This measurement gives the lever arm $\alpha=eV_{\rm SD}/\Delta V_P$ without additional measurements and delivers the temperatures $T_L$ and $T_R$ of each reservoir from a single $I_{\rm DC}(V_P)$ sweep. This allows a temperature measurement without calibration by another thermometer, thus constituting a primary thermometer. As an alternative, the differential conductance through the dot can be measured using a small AC voltage, resulting in the derivative of the FD function~\cite{vanhouten_1992}.

We note that here, the device is operated in a highly non-linear regime where the dot current $I_{\rm DC}$ depends only on the tunneling rate $\Gamma$ but is -- to lowest order -- independent of the applied bias $k_B T \ll V_{\rm SD}\ll \Delta$ once the dot level is well within the transport window spanned by source and drain chemical potentials. However, the electrons traversing the dot are injected at a high energy $V_{\rm SD}\gg k_B T$ into the reservoir with the lower chemical potential. These hot electrons will relax their energy and thereby cause heating in the 2DEG reservoir. The currents and biases used here are rather small, typically giving heating powers $\sim I_{\rm DC}V_{\rm SD}$ below $1\,$fW. Nevertheless, this heat will need to be removed, e.g. through the ohmic contacts or the phonon degree of freedom. We experimentally choose the bias $V_{\rm SD}$ small enough to avoid measurable heating.

For ultra-low temperatures, one critical aspect of the quantum dot thermometer is the requirement to have a dot level much sharper than the FD distribution to be probed and resolved. The broadening of the dot level is given by lifetime broadening: the finite time an electron spends on the dot, defined by its escape rate $\sim \Gamma$, introduces an uncertainty on its energy through the time-energy Heisenberg uncertainty principle. In gate defined dots, the tunneling rate $\Gamma$ can be tuned widely over many orders of magnitude with gate voltages, affording broad flexibility. While $\Gamma$ can easily be made sufficiently small to satisfy $h \Gamma\ll k_B T$ even at the lowest temperatures, reduced $\Gamma$ also suppresses the dot current $I_{\rm DC}\sim e \Gamma /2$. Taking $2h\Gamma= k_B T$, an upper bound on the dot current of $I\sim1\,$pA$\cdot \vartheta$ results, where $\vartheta$ is the temperature in mK. Thus, to be clearly in the temperature broadened regime, currents far below these upper bounds are required, setting a practical limit of order of $10\,$mK as the lowest temperature that can be measured with the current setup.

An integrated charge sensor directly adjacent to the quantum dot~\cite{field_1993,elzerman_2003} makes it possible to overcome this limitation: a measurement of the average dot charge occupation while sweeping the dot level through a charge transition~\cite{dicarlo_2004} reflects the FD distribution under similar conditions as described before. However, the dot-reservoir tunneling rate $\Gamma$ can now be made essentially arbitrarily small, ensuring $h \Gamma\ll k_B T$ even for temperatures well below $1\,$mK. This is possible because the size of the charge sensor signal is nearly independent of $\Gamma$ and the charge sensor remains operational for arbitrarily small $\Gamma$. The distribution function is conveniently measured when the dot tunneling is fast compared to the data acquisition rate, avoiding complications due to real time detection of single electron tunneling. The current through the charge sensor still gives rise to phonon or photon emission~\cite{gasparinetti_2012} and generally causes heating, analogous to a current flowing directly through the dot as discussed above. However, the sensor and its reservoirs can be electrically isolated and spatially separated somewhat from the dot, reducing heat leaks and coupling strength~\cite{granger_2012} and improving the situation compared to a direct current through the quantum dot. Nevertheless, the sensor biasing will need to be experimentally chosen to minimize such heating effects.

Similar thermometry can also be performed in a double quantum dot configuration, where charge transitions involving a reservoir can be used to measure the FD distribution and the corresponding temperature. The relevant double dot lever arm can be extracted again from finite bias measurements~\cite{vanderwiel_2002} or can be calibrated at elevated temperatures where it is safe to assume $T_{\rm MC} = T_{L,R}$. It is worth noting that in a double dot, the thermal smearing of the reservoirs can be essentially eliminated when studying internal transitions such as inter-dot tunneling, allowing measurements with a resolution much better than the reservoir temperature~\cite{vanderwiel_2002}. Nevertheless, internal double dot transitions can also be used for reservoir thermometry depending on the dot configuration~\cite{dicarlo_2004}. Similarly, in optically active semiconductor quantum dots, the reservoir temperature can be irrelevant, and the optical line width is limited by the lifetime and/or other noise sources such as semiconductor charge instabilities or nuclear spin noise~\cite{kuhlmann_2013}.

\begin{figure*}[*t]
\includegraphics[width=15cm]{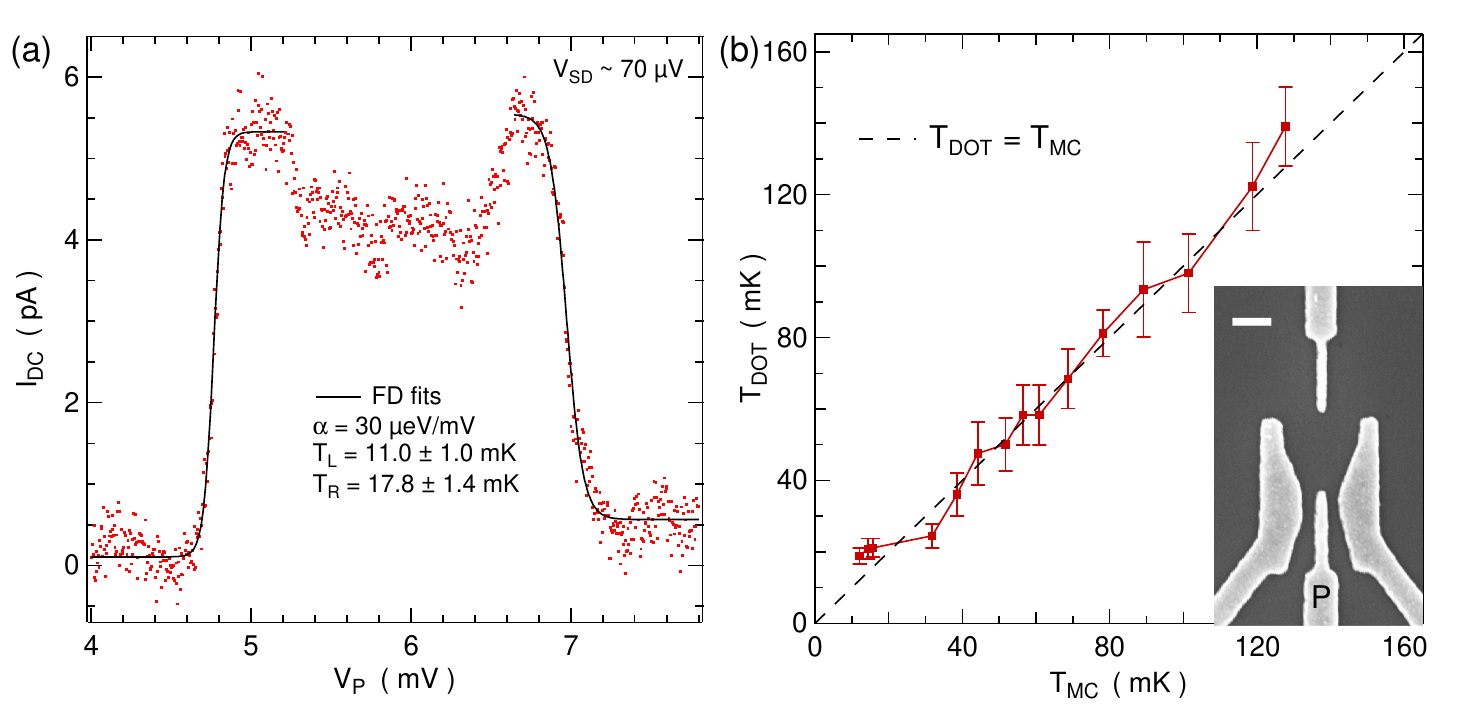}
\caption{(a)~DC current $I_{\rm DC}$ through a single quantum dot as a function of plunger gate voltage $V_P$ at refrigerator temperature $T_{\rm MC} = 9$~mK, showing a high current region (dot level between source and drain chemical potential) and a low current region (dot level outside source-drain window). These regions are separated by the Fermi-Dirac distributions in each reservoir, separately giving $T_L$ and $T_R$ from Fermi-Dirac fits. The right reservoir is connected to the current preamplifier and slightly warmer than the left reservoir. (b)~$T_L$ averaged over several repeated measurements as a function of $T_{\rm MC}$. The dot configuration was not changed during the temperature sweep. Inset: SEM picture of a device similar to the one measured (P: plunger gate, scale bar: 200~nm).}
\label{fig:1}
\end{figure*}

Interestingly, the energy levels of the double dot can easily be configured (e.g. sufficiently far away from the triple points or bias triangles) so that no net current can flow through the double dot even at some finite bias (here always assuming sequential tunneling only), avoiding dissipative heating originating from the double dot altogether. Despite the absence of current flow,  the system can still easily be probed with a charge sensor, and the reservoir temperature can be extracted as described above. A similar situation can also be exploited in a single dot with one barrier tuned to be very opaque \cite{AmashaT1}. The biasing of the charge sensor nevertheless still dissipates energy, as already described.

\section{Thermometry with direct transport}
\label{directtransport}

The quantum dots were fabricated with standard UV and ebeam lithography and evaporation of Ti/Au depletion gates. The single quantum dot (SQD) layout, see inset of Fig.\,\ref{fig:1}(b), was adapted from Ref.\,\cite{ciorga_2000}, giving access to the few electron regime in transport measurements. The 2DEG is formed at a single AlGaAs/GaAs interface, located 110\,nm below the surface, with charge carrier density $n = 2.8 \cdot 10^{11}\,\mbox{cm}^{-2}$ and mobility $\mu = 280'000\,\mbox{cm}^2/\mbox{(Vs)}$. This wafer was chosen because of excellent charge stability. The devices were cooled down without positive voltage bias on the gates. The ohmic contacts are non-magnetic, made from AuGe/Pt, and optimized for minimal contact resistances, typically $\sim100\,\mathrm{\Omega}$. The direct current $I_{\rm DC}$ through the dot was measured with a $3\,$Hz low-pass filter.

We now show how the reservoir temperatures $T_L$ and $T_R$ can be extracted from a measurement of the current $I_{\rm DC}$ through the dot at finite applied bias $V_{\rm SD}$ as a function of the plunger gate voltage $V_P$, as shown in Fig.\,\ref{fig:1}(a). The plunger gate $V_P$ allows us to shift the energy of the dot energy level through both source and drain chemical potentials without significantly changing the reservoir tunneling rates for a small change of $V_P$: more negative $V_P$ capacitively shifts the dot level to higher energy. A finite current flows through the dot when the dot energy level is located within the transport window. Otherwise, no current can flow, either due to a lack of filled electron states when the dot energy is above the higher chemical potential reservoir, or due to a lack of empty states the dot electron can tunnel into when the dot energy is below the lower chemical potential reservoir. The transitions between zero and finite current each reflect the distribution function of the respective reservoir, and can be fit by a FD function of the form

\begin{equation}
\label{eq:FD}
f_{FD}(V_P) = I_1 \left[\exp\left( \frac{\alpha (V_P - V_{P0})}{k_B T_{L,R}} \right) + 1\right]^{-1} +I_0 ,
\end{equation}

\noindent with step height $I_1$, offset current $I_0$ and plunger gate offset $V_{P0}$. Note that a rising (falling) step is obtained by the choice of the relative sign of $I_0$ and $I_1$. This generally gives high quality fits of the current steps, see Fig.\,\ref{fig:1}(a), and delivers separate temperatures $T_{L,R}$ for the left and right reservoirs, respectively. The reservoir connected to the current preamplifier gives slightly higher temperatures compared to the other reservoir, see Fig.\,\ref{fig:1}(a). The weak dependence of dot current on $V_P$ in the high current state can arise e.g. due to small variations in the local density of states in the leads, and is not further pursued here. The DC bias voltage was reduced until no effects on the extracted temperatures was observed, typically $V_{\rm SD} < 100\mbox{ }\mu$V at the lowest temperatures -- still allowing to clearly separate the two flanks.

The error-bars on the individual fits often turn out rather small ($\lesssim 10\,\%$), see Fig.\,\ref{fig:1}(a), yet a larger uncertainty becomes apparent when the fits are performed over a large number (of order 10) of repeated current traces under nominally identical conditions. This uncertainty is due to charge instabilities and resulting random telegraph noise -- occasionally directly identifiable in the data -- as well as slow drifts in the 2DEG material and quantum dots, or occasionally also external influences. We note that the sensitivity to such disturbances becomes more pronounced at lower temperature, already requiring an energy jitter of much less than $\sim 1\,\mathrm{\mu eV}$ at 10\,mK -- a quite remarkable charge stability \cite{kuhlmann_2013}. The severity of such charge noise depends sensitively on the detailed dot gate voltage configuration as well as the wafer material and fabrication procedure, and can become negligible at elevated temperatures due to increased thermal broadening. Current traces with obviously apparent switching events are not included in the ensemble of traces used to extract temperature.

In Figure~\ref{fig:1}(b), we repeat this procedure to extract the average temperature $T_{\rm DOT}$ measured with the quantum dot at fixed configuration for a number of refrigerator mixing chamber temperatures $T_{\rm MC}$, measured with a Cerium-Magnesium-Nitrate (CMN) thermometer. The CMN thermometer was calibrated using a standard fixed point device with 6 superconducting transitions between $1.2\,$K and $96\,$mK, giving excellent agreement between fixed point device and CMN. The average dot temperature is determined from several repeated current traces for each $T_{\rm MC}$, including only $T_L$ in the average. The resulting standard deviation is used to give the error bars on $T_{\rm DOT}$. As seen in Fig.\,\ref{fig:1}(b), we find decent agreement between $T_{\rm DOT}$ and $T_{\rm CMN}$ within the error bars over the temperature range from $\sim 20\,$mK to $\sim 130\,$mK, though at the lowest temperatures, $T_{\rm DOT}$ appears to saturate at $\sim 20\,$mK for the particular gate configuration used for this temperature sweep. When the measurement is further optimized and the tunnel rates are decreased a bit more (trading off current signal amplitude), the lowest temperature we extract in direct current through the dot is $T_{\rm DOT}=11\pm3$\,mK averaged over several traces, similar to the data shown in Fig.\,\ref{fig:1}(a), and within the error bars of the base temperature $T_{\rm MC}=9$\,mK.

\section{Thermometry with charge sensing}
\label{chargesensing}

\begin{figure*}
  \includegraphics[width=15cm]{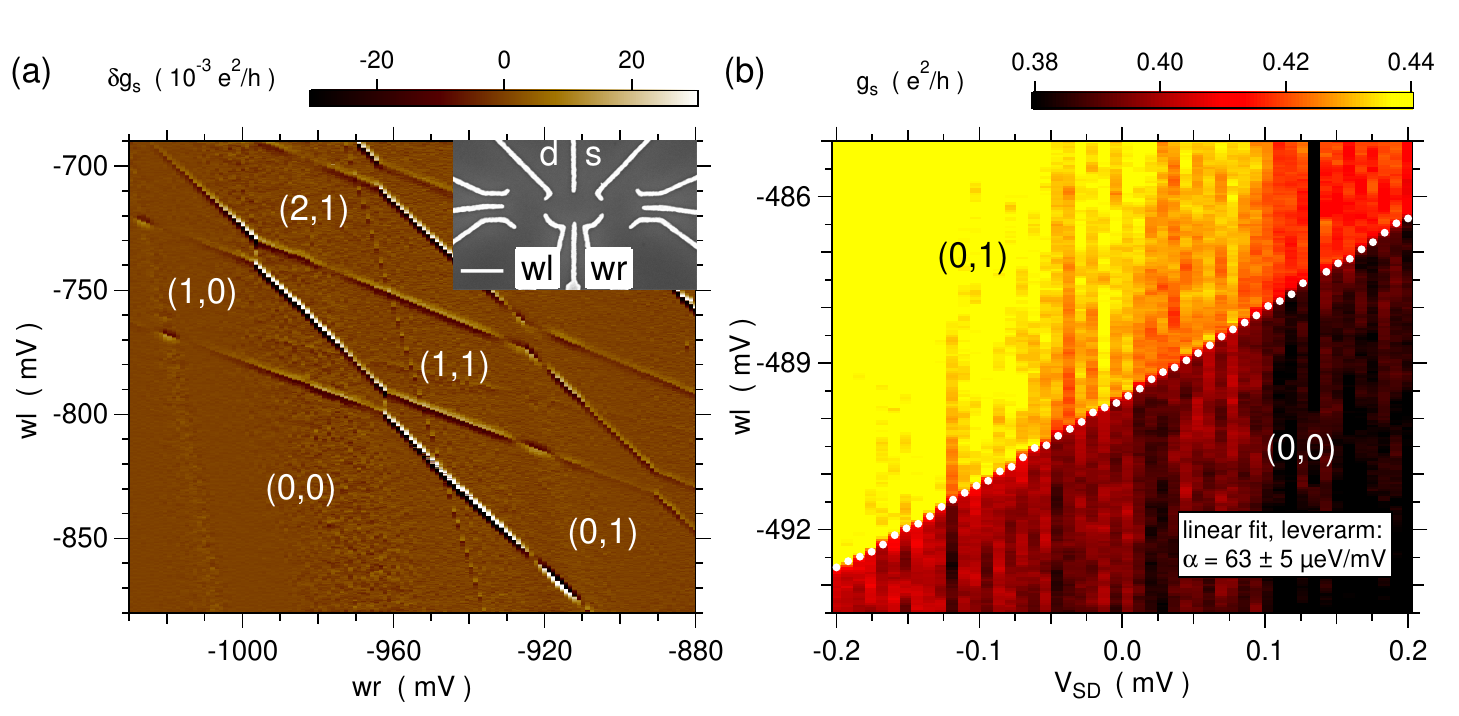}
\caption{(a)~Change in differential conductance $\delta g_s$ of the sensor on the right side measured as a function of the voltage on the left wall wl and right wall wr of the double dot. The average of each vertical trace was subtracted to improve visibility. The charge stability diagram shows the honeycomb structure typical of a double dot. The absolute electron occupation (n,m) is labeled, indicating the charge state in the left and right dot, respectively. Inset: SEM picture of a device similar to the one measured (d/s: drain/source, scale bar: 400~nm). (b)~Sensor differential conductance $g_s$ as a function of wl and $V_{\rm SD}$ around the (0,0) to (0,1) transition, allowing extraction of the lever arm $\alpha$, see text.}
\label{fig:2}
\end{figure*}

We now turn to thermometry with a charge sensor adjacent to a double quantum dot device. The design of the device was adapted from Ref.\,\cite{barthel_2010}, see inset of Fig.\,\ref{fig:2}(a), employing quantum dots as very sensitive charge detectors, directly adjacent on either side of the double dot. Here, we focus on data from one of the sensors since the other sensor gave very similar results. A GaAs 2DEG material very similar to the wafer used for the single dots was used, again experimentally tested to exhibit excellent charge stability. The differential conductance $g_s = dI/dV$ of the charge sensing quantum dot was measured with standard analog lock-in technique with an AC bias voltage $\leq 2\,\mathrm{\mu V}$. The sensor bias voltage was carefully experimentally restricted to avoid excess heating. The voltage and current noise of the measurement setup was carefully monitored and minimized, with optimal rms values of 0.5~$\mu$V and 50~fA, respectively.

The charge sensor was operated in the lifetime broadened regime, tuned on a steep slope of a Coulomb blockade peak, giving excellent signals of up to 100\% change of sensor conductance per electron. This is clearly superior to typical quantum point contact charge sensors, as previously reported~\cite{barthel_2010}. Even better sensitivities could be achieved when tuning the sensor dot into the temperature broadened regime, where much narrower, sharp peaks result. However, staying on such a sharp peak becomes experimentally difficult due to parasitic capacitive coupling between double dot gates and the sensor dot. Once the sensor is shifted into a Coulomb blockade valley or peak, where the slope is very small, the charge sensitivity is lost. Already in the lifetime broadened sensor regime used here, changes on the double dot gate voltages needed to be carefully compensated on the plunger gate of the sensor dot in order to maintain charge sensitivity.

The double dot charge stability diagram, as measured with the charge sensor, is shown in Fig.\,\ref{fig:2}(a) as a function of gate voltage on the left wall wl and right wall wr of the double dot, as labeled in the inset. The typical honeycomb pattern as expected for a double dot~\cite{vanderwiel_2002} is observed. Each dot can be emptied of all electrons (bottom left), as evidenced by the absence of further charge transition lines in the diagram at more negative gate voltages. This allows us to label the double dot charge state (n,m), indicating the absolute electron occupation in the left and right dot, respectively. A couple of additional charge transitions are also appearing with slopes deviating from those occurring in the honeycomb of the double dot, presumably due to some nearby charge traps in the semiconductor. These are also related to the charge noise as seen in the temperature measurements.

The reservoir temperature can again be extracted, here from the charge sensor signal with analogous FD fits to any of the charge transitions in the honeycomb involving one of the reservoirs. As before, the corresponding lever arm is required for the conversion from gate voltage to energy, and is extracted from measurements at high enough temperatures where double dot reservoir temperature $T_{S}$ measured with the sensor is equal to $T_{\rm MC}$. Bias triangles were not accessible in the regime the double dot was operated here due to tunnel rate asymmetries. We note that the inter-dot tunnel rate was tuned to be very small for the temperature measurements, with the double dot operated in a different gate voltage configuration as shown in Fig.\,\ref{fig:2}(a).

\begin{figure*}[*t]
\includegraphics[width=15cm]{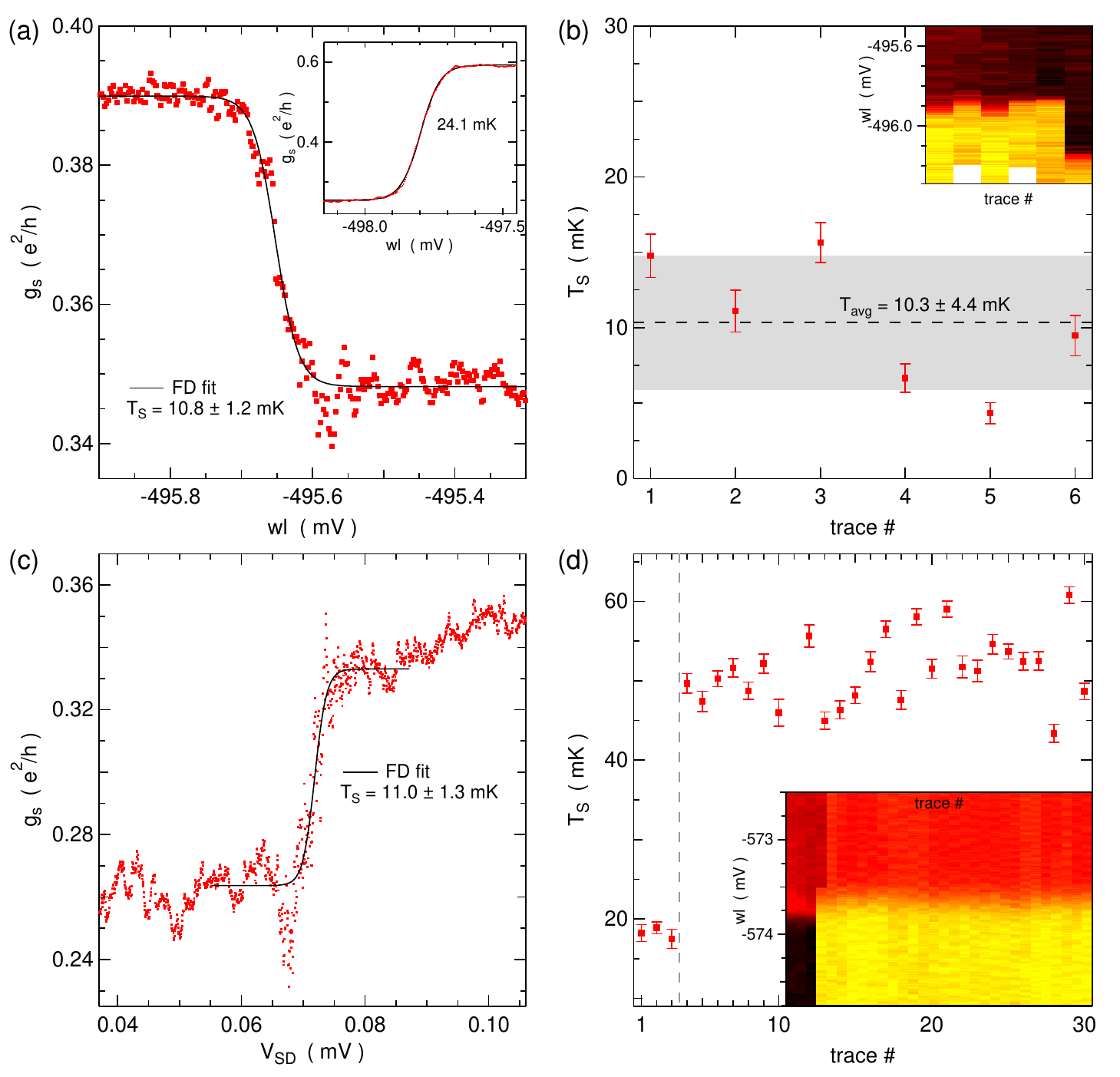}
\caption{(a)~Sensor differential conductance $g_s$ (sensor dot on the right side) as a function of gate voltage wl, showing the transition from the (0,0) to (0,1) charge state. The reservoir temperature $T_S$ is extracted from a FD fit (black curve) to sensor data, as indicated. Inset: Similar measurement at elevated temperature as labeled. (b)~Series of repeated $T_S$ measurements in the same dot configuration with an average temperature $T_{\rm avg}=10.3 \pm 4.4$~mK (dashed line: average; gray shaded area: standard deviation). Inset: Corresponding sensor conductance $g_s$ as a function of gate voltage wl versus trace number. (c)~Sensor conductance $g_s$ of the right charge sensor as a function of $V_{\rm SD}$ at the transition from (0,0) to (0,1), with FD fit (black curve) and extracted temperature (see text) as labeled. (d)~Reservoir temperature $T_S$ extracted with the sensor from several repeated wl sweeps (see inset) versus trace number, showing an abrupt change of the electronic dot configuration after three sweeps, which increases the temperature reading from 18~mK to 52~mK.}
\label{fig:3}
\end{figure*}

Alternatively, the same charge transition can be followed for various double dot source-drain voltages $V_{\rm SD}$ applied to the reservoir involved in the transition, as shown in Fig.\,\ref{fig:2}(b). Due to a finite capacitance of this reservoir to the dot, this gives an upper bound for the lever arm and the extracted temperature. However, the lever arm extracted at high temperature turns out to be the same as the upper bound (within the error bars of about 10\%), thus indicating that the reservoir-dot capacitance is small compared to the total dot capacitance for the configurations used in our device -- at least at very low tunnel rates utilized here. Hence, the slope of the charge transition line in the wl-$V_{\rm SD}$ plot gives the inverse of the lever arm. All temperature measurements shown here were carried out at the transition from (0,0) to (0,1), although similar results were obtained for other transitions.

Figure~\ref{fig:3}(a) shows a charge sensor measurement through the (0,0)-(0,1) transition and a FD fit at $T_{\rm MC} = 9$~mK, resulting in $T_{S} = 10.8 \pm 1.2$~mK. While the data gives very good agreement with the FD fits at slightly elevated temperatures~(see Fig.\,\ref{fig:3}(a) inset, giving $T_S=24.1\,$mK), the charge sensor temperature measurement again becomes more difficult at the lowest temperatures. The inset of Fig.\,\ref{fig:3}(b) shows the sensor signal for the same charge transition repeated a few times under identical conditions. Both the position and width of the transition is seen to fluctuate as a function of time, resulting in fluctuating temperatures $T_S$ extracted with the FD fit, see Fig.\,\ref{fig:3}(b), similar as described for temperature measurements via current through the dot. The error bars shown here (and also in Fig.\,\ref{fig:3}(d)) are from the FD fit only and do not include the uncertainty of about 10\% from the lever arm. In addition, the configuration of the sensor can also affect the extracted temperatures, typically resulting in lower temperatures for weaker sensor-double dot coupling, thus giving a smaller step height and making fitting more difficult. As before, curves displaying obvious switching events are not considered for determining temperature.

We can also use the double dot source-drain voltage $V_{\rm SD}$ instead of gate voltage to drive the charge transition and directly obtain a temperature value without needing a lever arm, since the reservoir-dot capacitance is small here, as previously discussed. In this way, we obtain an upper bound on the reservoir temperature which here is close (within 10\%) to the actual temperature. Such a $V_{\rm SD}$ charge transition measurement is illustrated in Fig.\,\ref{fig:3}(c), again for the (0,0)-(0,1) transition, and gives a very similar temperature as obtained from the gate sweep. The overshoot before and after the transition has been observed in several measurement curves at the lowest temperatures, both by sweeping $V_{\rm SD}$ or a gate, though it is not seen in some other traces, e.g. Fig.\,\ref{fig:3}(a). The origin of the overshoot is not currently understood.

The extreme sensitivity of the charge transition to the electrostatic environment is demonstrated in Fig.\,\ref{fig:3}(d). While scanning the same transition 30 times, an abrupt change in the charge configuration during the fourth scan has changed the charge sensor response considerably, even inverting the sign of the sensor response to the dot charge transition. This switching event caused the apparent FD fit temperature to change from 18~mK to 52~mK, while it is likely that the actual reservoir temperature was not affected. Scanning charge transitions different from (0,0) to (0,1) revealed similar temperatures but also suffered from the same problems with charge instabilities.

\section{Discussion}
\label{Discussion}

After considerable experimental efforts due to the pronounced sensitivity to electronic noise and device charge instabilities, we approach mixing chamber base temperature with both methods, direct transport and charge sensing. By using the nuclear refrigerator ($T_{\rm NR} < 1$~mK~\cite{casparis_2012}), no further reduction of the electron temperature was observed. In the direct transport measurements, we might suspect lifetime broadening of the quantum dot level as a limiting factor. But the temperatures obtained with the charge sensor are not evidently lower than the temperatures measured in direct transport, despite much lower dot tunneling rates.

In direct transport, dissipative heating from the voltage drop over the dot will eventually become significant at sufficiently low $T$. Estimates of the electron temperature $T$ assuming dominant Wiedemann-Franz cooling, an ohmic contact resistance of $100\,\rm\Omega$, $V_{\rm SD}=100\,\mathrm{\mu V}$ and a current of $8\,\mathrm{pA}$ ($\Gamma/2 = 50\,$MHz) indicate that the temperature is only increased by $\Delta T = 0.3\,$mK above the bath temperature at $T_{\rm MC} = 10\,$mK. At a much lower temperature $T_{\rm NR}=1\,$mK, however, the electron temperature is estimated to rise to $T = 2.8\,$mK due to poor thermal contact. This strong increase is due to the ohmic contact resistance, which could potentially be further reduced with improved fabrication. In addition, the voltage bias $V_{\rm SD}$ can also still be made much smaller, since a temperature of $T_{\rm NR}=1\,$mK corresponds to a broadening of the FD distribution of only $\sim 0.1\,\mathrm{\mu V}$, thus still leaving room to fulfill $e V_{\rm SD}\gg k_B T$.

Our experiments indicate that the electronic noise and external disturbances in the measurements setup play a very important role: excess voltage noise clearly increases the temperatures extracted. Filtering and shielding can be further improved, though already in the present experiment, a significant amount of work was invested \cite{casparis_2012}. We obtain noise levels as low as several hundred nanovolts across the dot measured at room temperature, but significantly less at the cold device due to filtering. The electron temperature becomes independent of the noise power in the lowest noise range, indicating that electronic noise is not the only or not the dominating limitation. The role of the charge sensor as a noise source and possible effects of coupling, back action~\cite{granger_2012} or sensor heating require further investigation.

The devices used here have outstanding charge stability, with noise on the dot energy level well below $1\,\mathrm{\mu eV}$~\cite{kuhlmann_2013,dial_2013}, making possible temperature measurements as low as $\sim10\,$mK presented here. Still, device charge instabilities present a serious obstacle if much lower temperatures are to be reached, already severely impeding the measurements here. The temperature measurement would benefit from faster measurements, thus cutting of the noise spectrum at the lowest frequencies and reducing the effect of random telegraph noise. We emphasize that the charge switching noise exceeds other setup noise such as the voltage sources on the gates, preamplifiers and Johnson noise of the sample wires.

Besides semiconductor charge instabilities, the GaAs nuclear spins can also act as a noise source, giving rise to a fluctuating Zeeman splitting and thus broadening of the single electron energy level (though the energy of a spin singlet would be immune to this noise). With GaAs hyperfine coupling constant $A=90\,\mathrm{\mu eV}$ \cite{paget_1977} and number of nuclear spins $N\sim10^5$ to $10^6$ enclosed in the electron wave function \cite{coish_2004}, the resulting energy fluctuations are of order $A/\sqrt{N}\sim0.1\,\mathrm{\mu eV}$, and become a limiting factor for $T\lesssim 1\,$mK. Finally, heat release from sample holder or other components can also be a limiting factor, resulting in temperatures decaying slowly over a timescale of days.

In conclusion, we have measured the reservoir electron temperature $T$ with a GaAs quantum dot in both direct transport and charge sensing. We find decent agreement with a CMN thermometer over a broad temperature range and we have demonstrated temperatures as low as $10\,\pm3$mK, reaching the state of the art in GaAs quantum dot thermometry \cite{potok_2007}. Currently, the main limitations are charge switching noise in the GaAs device, external electronic noise, heating effects due to the charge sensor as well as heat release. Lower temperatures might be achievable by further improving the setup and device, e.g. by better shielding and filtering, choosing materials with lower heat release and possibly optimizing the wafer material and device fabrication.

\begin{acknowledgements}
We would like to thank G.~Frossati, G.~Pickett, V.~Shvarts, P.~Skyba, M.~Steinacher and A.~de~Waard for valuable inputs.
This work was supported by Swiss Nanoscience Institute~(SNI), NCCR QSIT, Swiss NSF, ERC starting grant, and EU-FP7 MICROKELVIN and SOLID.
\end{acknowledgements}

\bibliographystyle{spphys}       

%
%

\end{document}